\documentstyle[preprint,prc,eqsecnum,aps]{revtex}
\begin{document}
\draft
\title{Shell model Monte Carlo calculations\\
for ${}^{170}$Dy}
\author{D.~J.~Dean, S.~E.~Koonin, G.~H.~Lang, P. B. Radha, and
W.~E.~Ormand}
\address{W. K. Kellogg Radiation Laboratory,
California Institute of Technology,\\
Pasadena, CA 91125 USA}
\date{\today}
\maketitle

\begin{abstract}
We present the first auxiliary field Monte Carlo calculations for a
rare earth nucleus, ${}^{170}$Dy. A pairing plus quadrupole
Hamiltonian is used to demonstrate the physical properties that can
be studied in this region. We calculate various static observables
for both uncranked and cranked systems and show how the shape
distribution evolves with temperature. We also introduce a
discretization of the path integral that allows a more efficient
Monte Carlo sampling.
\end{abstract}

\section{Introduction}

Although mean-field models of rare-earth nuclei can describe the
gross interplay between collective and single-particle degrees of
freedom \cite{good}, it is of obvious interest to pursue a fully
microscopic description to the greatest extent possible, particularly
as the angular momentum and/or temperature is increased. As we have
previously developed and demonstrated Monte Carlo methods for
treating very large basis shell model calculations
\cite{calvin,gladys}, it is natural to attempt to extend them to the
rare-earth region. Toward this end, we present here the first
auxiliary field Monte Carlo calculations of observables in a
rare-earth nucleus. For demonstration purposes, we have chosen the
pairing plus quadrupole Hamiltonian and consider the mid-shell
nucleus ${}^{170}$Dy.

As we move up to the rare earth region from the $sd$- and $fp$-shell
spaces, new physical and computational problems must be solved. The
neutrons and protons occupy different major shells, and so isospin
symmetry is lost. Further, the numbers of single-particle orbitals
and valence particles are larger. Most importantly, the long
autocorrelations times encountered in Metropolis sampling
\cite{gladys,ref8} of continuous auxiliary fields have to be avoided
by a novel discretization of the fields based on gaussian quadrature.
The small level spacing of these larger nuclei also requires
calculations at lower temperatures.

Our presentation is organized as follows. In Section II we review the
auxiliary field Monte Carlo method, discuss our choice of the
Hamiltonian, and introduce a discretization of the path integral to
facilitate the Monte Carlo sampling. We present results of static
observables and deformation distributions for ${}^{170}$Dy at various
temperatures and cranking frequencies in Section III, and draw
several conclusions about future directions.

\section{Formalism}

Details of the auxiliary field Monte Carlo approach to the shell
model have been presented elsewhere \cite{calvin,gladys}, so that we
only outline the method here. Given some many-body Hamiltonian $H$,
we desire a tractable expression for the imaginary time evolution
operator,
\begin{equation}
U=\exp\left(-\beta H\right)\;,
\end{equation}
where $\beta$ has units of inverse energy (MeV$^{-1}$, where
$\hbar=1$) and $\beta^{-1}$ can be interpreted as the temperature of
the system. The operator $H$ is a generalized many-body Hamiltonian,
and may contain terms such as $-\mu N$ in the grand-canonical
ensemble, or $-\omega J_z$ in cranked systems. (In these cases, $\mu$
is the chemical potential and $\omega$ is the cranking frequency.)
The partition function is
\begin{equation}
Z={\rm Tr}\exp\left(-\beta H\right)\;,
\end{equation}
from which we can construct the thermal expectation value of an
operator $O$ as
\begin{equation}
\langle O\rangle=
\frac{1}{Z}{\rm Tr}
\left[O\exp\left(-\beta H\right)\right]\;.
\end{equation}
Here, Tr is the trace over many-body states of fixed (canonical) or
all (grand-canonical) particle number.

We restrict ourselves to generalized Hamiltonians that contain at
most two-body terms. The Hamiltonian can then be written as a
quadratic form in some set of convenient operators $\{{\cal
O}_\alpha\}$,
\begin{equation}
H=\sum_\alpha \varepsilon_\alpha {\cal O}_\alpha+
\frac{1}{2} \sum_\alpha V_\alpha {\cal O}_\alpha^2\;,
\label{quadr}
\end{equation}
where we have written the quadratic term in diagonal form. These
operators are typically either one-particle (density) or
one-quasiparticle (pairing), and the strength of the two-body
interaction is characterized by the real numbers $V_\alpha$ related
to the two-body matrix elements.

For $H$ in the quadratic form (\ref{quadr}), we can write the
evolution operator as a path integral. The exponential is first split
into $N_t$ `time' slices, $\beta=N_t\Delta\beta$, so that
\begin{equation}
U=\left[\exp \left(-\Delta\beta H\right) \right]^{N_t}\;.
\end{equation}
A Hubbard-Stratonovich (HS) \cite{hs} transformation is then
performed on the two-body term in each time slice $n=1,\ldots,N_t$
yielding for the evolution operator
\begin{equation}
U=\left[\exp\left(-\Delta\beta H\right) \right]^{N_t}
\simeq
\int{\cal D}[\sigma]G(\sigma)U_\sigma\;,
\end{equation}
where the integration measure is
\begin{equation}
{\cal D}[\sigma]=
\prod^{N_t}_{n=1} \prod_\alpha d\sigma_{\alpha n}
\left( \frac{\Delta\beta\mid V_\alpha\mid}{2\pi}
\right)^{\frac{1}{2}}\;,
\end{equation}
the Gaussian factor is
\begin{equation}
G(\sigma)=
\exp\left(-\sum_{\alpha n} \frac{1}{2} \Delta\beta\mid V_\alpha\mid
\sigma^2_{\alpha n}\right)\;,
\label{gaussian}
\end{equation}
the one-body evolution operator is
\begin{equation}
U_\sigma\equiv U_{N_t} U_{N_{t-1}} \ldots U_1\;,
\end{equation}
with $U_n\equiv \exp \left[-\Delta\beta h_\sigma(\tau_n)\right]$, and
the one-body Hamiltonian is
\begin{equation}
h_\sigma(\tau_n)=
\sum_\alpha \left(\varepsilon_\alpha+ s_\alpha V_\alpha
\sigma_{\alpha n}\right) {\cal O}_\alpha\;,
\end{equation}
where the phase factor $s_\alpha$ is $\pm 1$ if $V_\alpha<0$, and is
$\pm i$ if $V_\alpha > 0$. Each real variable $\sigma_{\alpha n}$ is
the auxiliary field associated with ${\cal O}_\alpha$ at the time
slice $n$.

Expectation values of observables are calculated through
\begin{equation}
\langle{\cal O}\rangle=
\frac{\int {\cal D}[\sigma] {\cal G}(\sigma) \zeta(\sigma)
\langle{\cal O}\rangle_\sigma}
{\int{\cal D}[\sigma] {\cal G}(\sigma) \zeta(\sigma)}\;,
\label{expval}
\end{equation}
where
\begin{equation}
\zeta(\sigma) = {\rm Tr}\;U_\sigma
\end{equation}
is the partition function for each field configuration and
\begin{equation}
\langle{\cal O}\rangle_\sigma=
\frac{{\rm Tr}\left[{\cal O} U_\sigma\right]}
{\zeta(\sigma)}\;.
\end{equation}

To perform the integration via Monte Carlo methods, we introduce the
non-negative integration weight $W(\sigma)={\cal
G}(\sigma)\mid\zeta(\sigma)\mid$, and the ``sign''
$\Phi(\sigma)=\zeta(\sigma)/\mid\zeta(\sigma)\mid$, so that the
expression (\ref{expval}) for an observable becomes
\begin{equation}
\langle{\cal O}\rangle=
\frac{\int{\cal D}[\sigma] W(\sigma) \Phi(\sigma)
\langle{\cal O}\rangle_\sigma}
{\int{\cal D}[\sigma]W(\sigma) \Phi(\sigma)}\;.
\label{biginteg}
\end{equation}

The integral (\ref{biginteg}) can be evaluated by Monte Carlo methods
using samples generated by the algorithm of Metropolis {\it et al.}
\cite{ref8}, as described in Ref.~\cite{gladys}. However, in view of
the large number of auxiliary fields involved (some $10^5$), the
successive field configurations are highly correlated for a
reasonable acceptance fraction ($\sim0.5$), the autocorrelation
length being over 200 sweeps. To generate uncorrelated samples more
efficiently, we have approximated the continuous integral over each
$\sigma_{\alpha n}$ by a discrete sum derived from a Gaussian
quadrature. In particular, the relation
\begin{equation}
e^{\Delta\beta V{\cal O}^2/2}\approx
\int^\infty_{-\infty} d\sigma f(\sigma)
e^{\Delta\beta V\sigma{\cal O}}
\end{equation}
is satisfied through terms in $(\Delta\beta)^2$ if $f(\sigma)=
{1\over6} \left[\delta(\sigma-\sigma_0)+ \delta(\sigma+\sigma_0)+
4\delta(\sigma)\right]$, where {$\sigma_0= (3/V\Delta\beta)^{1/2}$}.
(Note that commutator terms render the HS transformation accurate
only through order $\Delta\beta$ anyway.)

In this way, each $\sigma_{\alpha n}$ becomes a 3-state variable and
the path integral becomes a (very large) path sum, which can be
sampled using the algorithm of Metropolis {\it et al.} We have found
that this discretization reduces the correlation length to only five
sweeps, and thus increases our efficiency by a factor of 40 relative
to the continuous case, with no loss of accuracy.

To describe rare-earth nuclei, we choose the $Z=50$--82 shell for
protons ($2s1d0g_{7/2}0h_{11/2}$) and the $N=82$--126 shell for
neutrons ($2p1f0g_{9/2}0i_{13/2}$). The Hamiltonian we have chosen is
of the pairing plus quadrupole form given by
\begin{equation}
H=\sum_\alpha\varepsilon_\alpha
a^\dagger_\alpha a_\alpha -g_p P^\dagger_p P_p-
g_n P^\dagger_n P_n- {\chi\over2} Q_p\cdot Q_n\;,
\end{equation}
where $\varepsilon_\alpha$ are the single particle energies,
$a^\dagger_\alpha$ and $a_\alpha$ are the anti-commuting creation and
annihilation operators associated with the single particle state
$\alpha$, $P^\dagger_{p(n)},P_{p(n)}$ are the monopole pair creation
and annihilation operators for protons and neutrons, and $Q_{p(n)}$
is the quadrupole-moment operator. The single particle states
$\alpha$ are defined by the complete set of quantum numbers
$nljmt_z$, denoting the principal, orbital angular momentum, total
single-particle angular momentum, $z$-projection of $j$, and the
third component of the isospin quantum numbers, respectively.
Single-particle operators are thus represented by matrices of
dimension 32 and 44, for protons and neutrons, respectively. The
pairing strengths $g_{p(n)}$, the quadrupole interaction strength
$\chi$, and the single particle energies used in our calculations are
given in Table~\ref{table1} (taken from Ref.~\cite{kumar}).

\section{The Calculation and Results}

To demonstrate the power of our methods, we study the mid-shell
nucleus ${}^{170}$Dy (16 valence protons and 22 valence neutrons),
which requires some $10^{21}$ $m$-scheme determinants. We have used
$\Delta\beta=0.0625$ and $N_t=8$ to 64 time slices.

In addition to grand-canonical calculations, we performed canonical
analyses of the fields generated through grand-canonical sampling. In
particular, canonical observables (subscript ``$c$'') are given in
terms of the grand-canonical sampling (subscript ``$g$'') as
\begin{equation}
\langle{\cal O}\rangle_c=
\frac{\int{\cal D}[\sigma] W_g(\sigma) \Phi_c(\sigma)
\left[\zeta_c(\sigma)/ \zeta_g(\sigma)\right]
\langle{\cal O}\rangle_{\sigma c}}
{\int{\cal D}[\sigma] W_g(\sigma) \Phi_c(\sigma)
\left[\zeta_c(\sigma)/ \zeta_g(\sigma)\right]}\;.
\end{equation}
The fluctuations in $\left[\zeta_c(\sigma)/\zeta_g(\sigma)\right]$
determine how precise this evaluation can be. We find that the
fluctuations are less than 10\% (as the number distribution in the
grand-canonical ensemble is small), so that canonical observables can
be calculated with good precision.

In Fig.~(\ref{beta}) we show the static observables for the uncranked
system in both the grand-canonical and canonical formalisms. We
calculated observables canonically, using grand canonical fields, up
to $\beta=2.0$ in order to demonstrate that for these nuclear
systems, either method may be used in this kind of calculation. Note
that as the temperature decreases, the nucleus becomes more deformed.
Relaxation of the expectation values of $H$ and $J^2$ is also clearly
seen. The sign in these cases is identically one for all
temperatures.

Cranking calculations in which $H\rightarrow H-\omega J_z$ have also
been performed. The systematics are shown in Fig.~\ref{crank}, where
we display
$\langle\Phi\rangle$,
$\langle H\rangle$,
$\langle Q^2\rangle$,
$\langle J^2\rangle$,
$-g\langle P^\dagger P\rangle$, and the moment of inertia,
$\langle{\cal I}_2\rangle$.
Note that the sign degrades quite rapidly with increasing $\omega$,
making cranking calculations at lower temperatures difficult. Moments
of inertia were calculated from ${\cal I}_2=d\langle
J_z\rangle/d\omega= \beta[\langle J^2_z\rangle-\langle
J_z\rangle^2]$. At high temperatures, the nucleus is unpaired and the
moment of inertia decreases as the system is cranked. However, for
lower temperatures when the nucleons are paired, the moment of
inertia initially increases as we begin to crank, but then decreases
at larger cranking frequencies as pairs break; Figure~2 shows that
the pairing gap also decreases as a function of $\omega$. It is well
known that the moment of inertia depends on the pairing gap
\cite{preston}, and that initially ${\cal I}_2$ should increase with
increasing $\omega$. Once the pairs have been broken, the moment of
inertia decreases. These features are evident in the figure.

In addition to static observables, we have calculated the nuclear
shape distributions. These distributions give a clear description of
various shape and phase transition phenomena \cite{yoram}. To obtain
a detailed picture of the deformation, we use the components of the
quadrupole operator $Q_\mu=r^2Y^{*}_{2\mu}$. Rotational invariance of
the Hamiltonian implies that the expectation value of each component
of $Q_\mu$ vanishes on average; however, each Monte Carlo sample will
have finite $Q_\mu$ values, from which we construct the quadrupole
tensor $Q_{ij}= 3x_ix_j-\delta_{ij}r^2$. The eigenvalues of the
latter tensor then lead directly to the deformation parameters
\cite{ormand93}.

Figure~\ref{bgp} shows the evolution of the shape distribution for
${}^{170}$Dy at inverse temperatures $T^{-1}=0.5$, 1.0, 2.0, and
3.0~MeV$^{-1}$. These contour plots show the free energy
$F(\beta,\gamma)$, obtained from the shape probability distribution,
$P(\beta,\gamma)$, by
\[
F(\beta,\gamma)= -T\ln
\frac{{\cal F}(\beta,\gamma)}{\beta^3\sin 3\gamma}\;,
\]
where the $\beta^3\sin3\gamma$ is the metric in the usual deformation
coordinates. As is seen from the plots, deformation clearly sets in
with decreasing temperature. At high temperatures, the system is
nearly spherical, whereas at lower temperatures, especially at
$T^{-1}=3.0$, there is a prolate minimum on the $\gamma=0$ axis.

These calculations demonstrate how auxiliary field Monte Carlo
methods can be extended to rare earth nuclei. For the first time we
have used different major shells for protons and neutrons in such a
calculation. We have demonstrated how to obtain canonical information
from grand-canonical sampling, and have introduced a discretization
of the field integrals that allows for a much more efficient sampling
of the integrand. Results for ${}^{170}$Dy with a pairing +
quadrupole Hamiltonian show qualitatively the expected behavior as a
function of both cranking frequency and temperature. While the
schematic nature of the Hamiltonian precludes reading too much into
our results, such calculations could be used to check various
approximation schemes. Furthermore, the recently demonstrated ability
to handle more realistic shell-model hamiltonians via Monte Carlo
methods \cite{alhassid} should allow these calculations to move
beyond the schematic level.

\acknowledgements
This work was supported in part by the National Science Foundation
(Grants No. PHY90-13248 and PHY91-15574) and by a Caltech DuBridge
postdoctoral fellowship to W.~E.~Ormand. We also acknowledge useful
discussions with Y.~Alhassid. Calculations were performed on the
Intel Touchstone Delta system operated by Caltech for the Concurrent
Supercomputer Consortium.

\begin{figure}
\caption{Canonical and grand-canonical observables for the uncranked
${}^{170}$Dy system as a function of $\beta$. Circles represent
results of canonical projection of the grand-canonical fields, while
squares show the grand-canonical results. Lines are drawn to guide
the eye and all error bars are smaller than the symbol sizes. We show
expectation values of the energy $\langle H\rangle$, the isoscalar
quadrupole moment $\langle Q^2\rangle$, the valence nucleon number
$\langle N\rangle$ and number variance $\langle\Delta
N\rangle=\protect\sqrt{\langle N^2\rangle-\langle N\rangle^2}$
(grand-canonical only), the squared angular momentum $\langle
J^2\rangle$, and $-g\langle P^\dagger P\rangle$, the expectation
value of the pairing terms in the hamiltonian.}
\label{beta}
\end{figure}

\begin{figure}
\caption{Grand-canonical observables for ${}^{170}$Dy at various
cranking frequencies and temperatures. We show the average sign
$\langle\Phi\rangle$, the isoscalar quadrupole moment $\langle
Q^2\rangle$, the energy $\langle H\rangle$, the square of the angular
momentum $\langle J^2\rangle$, the moment of inertia $\langle {\cal
I}_2\rangle$, and the expectation value of the pairing terms in the
hamiltonian $-g\langle P^\dagger P\rangle$. Error bars where not
shown are approximately the size of the symbols, and lines are drawn
to guide the eye.}
\label{crank}
\end{figure}

\begin{figure}
\caption{Contours of the free energy (as described in the text) in
the polar-coordinate $\beta$-$\gamma$ plane for the ${}^{170}$Dy
system. Inverse temperatures are from 0.5 (top) 1.0, 2.0, and
3.0~MeV$^{-1}$ (bottom). Contours are shown at 0.3~MeV intervals,
with lighter shades indicating the more probable nuclear shapes. As
our calculations become indeterminate at $\gamma=0$, we have
truncated these plots at small $\gamma$.
\label{bgp}}
\end{figure}

\begin{table}
\caption{Physical parameters used in these calculations.}
\label{table1}
\begin{tabular}{ldcld}
\multicolumn{2}{c}{Protons}&
&\multicolumn{2}{c}{Neutrons}\\
\multicolumn{1}{c}{Orbital}& $\epsilon$~(MeV)
&&\multicolumn{1}{c}{Orbital}& $\epsilon$~(MeV)\\
\tableline
$0g_{7/2}$ & $-$2.24 && $0g_{9/2}$ & $-$3.540\\
$1d_{5/2}$ & $-$1.979 && $1f_{7/2}$ & $-$3.211\\
$0h_{11/2}$ & $-$1.120 && $0i_{13/2}$ & $-$2.240\\
$1d_{3/2}$ & $-$0.122 && $2p_{3/2}$ & $-$1.20\\
$2s_{1/2}$ & 0.000 && $1f_{5/2}$ & $-$0.933\\
&&& $2p_{1/2}$ & 0.00\\
\multicolumn{2}{c}{$g_p=+0.160$ MeV}&&
\multicolumn{2}{c}{$g_n=+0.131$ MeV}\\
&& $\chi=+0.054$ MeV~fm$^4$\\
\end{tabular}
\end{table}

\end{document}